\documentclass[conference]{IEEEtran}

\usepackage[T1]{fontenc}
\usepackage{amsmath}
\usepackage{amssymb}
\usepackage{subfigure}
\usepackage{cite}
\usepackage{epsfig}
\usepackage{latexsym}
\usepackage{makeidx}
\usepackage{ifthen}
\usepackage{color}
\usepackage{multirow}
\usepackage{url}
\usepackage{graphicx}
\usepackage{algorithm}
\usepackage{algpseudocode}
\usepackage{balance}
\usepackage{times}
\usepackage{bussproofs}
\usepackage{pifont}
\usepackage{soul}
\usepackage{tabu}
\usepackage{array}
\usepackage{flushend}

\newcommand{\server}{${\cal S}$}

\mathchardef\mhyphen="2D

\newcommand{\dofac}[0]{{\normalsize \textsf{SOM}}}
\newcommand{\comrade}[0]{{\normalsize \textsf{Commune}}}

\newcommand{\F}[0]{\mathrm{F}}
\newcommand{\aont}[0]{$\mathrm{AON\mhyphen FFT}$}
\newcommand{\ant}[0]{\mathrm{AON\mhyphen FFT}}

\algrenewcommand\algorithmicindent{0.5em}

\newtheorem{definition}{Definition}

\newtheorem{observation}{Observation}

\newcommand{\rpermit}{\ensuremath{\mathrm{read}}}
\newcommand{\wpermit}{\ensuremath{\mathrm{write}}}

\begin{document}
\title{Commune: Shared Ownership in an Agnostic Cloud}

\author{
\IEEEauthorblockN{Claudio Soriente}
\IEEEauthorblockA{ETH Zurich}
\and
\IEEEauthorblockN{Ghassan Karame}
\IEEEauthorblockA{NEC Laboratories Europe}
\and
\IEEEauthorblockN{Hubert Ritzdorf}
\IEEEauthorblockA{ETH Zurich}
\and
\IEEEauthorblockN{Srdjan Marinovic}
\IEEEauthorblockA{ETH Zurich}
\and
\IEEEauthorblockN{Srdjan \v{C}apkun}
\IEEEauthorblockA{ETH Zurich}
}

\maketitle

\begin{abstract}
Although cloud storage platforms promise a convenient way for users to
share files and engage in collaborations, they require all files to
have a single owner who unilaterally makes access control decisions.
Existing clouds are, thus, agnostic to shared ownership.
This can be a significant limitation in many collaborations because
one owner can, for example, delete files and revoke access without consulting the other collaborators.

In this paper, we first formally define a notion of \emph{shared}
ownership within a file access control model. We then propose a
solution, called \textsf{Commune,} to the problem of distributively enforcing shared ownership
in agnostic clouds, so that access grants require the support of a pre-arranged threshold of owners.
\textsf{Commune} can be used in existing clouds without requiring any modifications to the platforms.
We analyze the security of our solution and evaluate its scalability and performance by means of an implementation
integrated with Amazon S3.
\end{abstract}

\section{Introduction}\label{sec:intro}
Even though the cloud promises a convenient way for users to share
files and effortlessly engage in collaborations, it still retains the
notion of \emph{individual} file ownership. That is, each file stored in the cloud is owned by a single
user, who can \emph{unilaterally} decide whether to grant or deny any access request to that file.
However, the individual ownership is not suitable for numerous cloud-based applications and collaborations.
Consider a scenario where a number of research organizations and
industrial partners want to set up a shared repository in the cloud to collaborate on a
joint research project.
If all participants contribute their research
efforts to the project, then they may want to share the ownership over
the collaboration files so that all access decisions are
agreed upon among the owners. There are two main arguments why this may be preferred to individual ownership. First, if there is
a sole owner, then he can abuse his rights by unilaterally making access control decisions.
The community features a number of anecdotes where malicious users revoke access to shared
files from other collaborators. This problem is further exacerbated by
users increasingly storing most of their data in the
cloud without keeping local copies,
and accessing them through portable devices that have limited storage capacity.
Second, even if owners are willing to elect and trust one of them to
make access control decisions, the elected owner may not want to be held
accountable for collecting and correctly evaluating other
owners' policies. For example, incorrect evaluations may incur
negative reputation or financial penalties.

In contrast to individual ownership, we introduce a novel
notion of \emph{shared ownership} where $n$ users jointly own a file
and each file access request must be granted by a pre-arranged threshold of $t$
owners. We remark that existing cloud platforms, such as Amazon S3 and
Dropbox, provide no support for shared ownership policies, and offer only
basic access control lists. In short, they are \emph{agnostic} to the
concept of shared ownership.
Furthermore, state-of-the-art trust management systems that can support
shared ownership policies (e.g., SecPAL \cite{SecPAL}, KeyNote \cite{KeyNote}, Delegation Logic \cite{DL})
make all access decisions using a \emph{centralized} Policy Decision Point (PDP).
This is not suitable for enforcing our shared ownership model, because
the user who administrates the PDP can arbitrarily change the policy rules set by the owners and enforce his own policies.

In this paper, we address the problem of \emph{distributed enforcement
of shared ownership within an agnostic cloud.}
By distributed enforcement, we mean enforcement where access to files in a shared repository is
granted if and only if $t$ out of $n$ owners separately support the grant
decision.
To tackle this problem, we first introduce the Shared-Ownership file
access control Model (\dofac{}) to define our
notion of shared ownership, and to formally state the
given enforcement problem.
We then propose our solution, called \comrade{}, that enforces
shared ownership policies in a \emph{distributed} fashion.
Additionally, \comrade{} can be used within a
third-party cloud without any modifications to the platform.
It only requires that the cloud offers
basic access control lists, as is the case with current platforms.
We
integrate a prototype implementation of \comrade{} within Amazon
S3~\cite{AmazonS} and we show that its performance scales well with the
file size and with the number of users.
To the best of our knowledge, \comrade\ is the first solution to distributed enforcement
of shared ownership in an agnostic cloud.

We summarize our contributions as follows:
\begin{itemize}

\item We formalize the notion of shared ownership within
a file access control model named \dofac, and use it to define a novel
access control problem of distributed enforcement of shared ownership
in an agnostic cloud.

\item We propose a solution, called \comrade{}, which distributively enforces \dofac{} and can be deployed in an agnostic
    cloud platform. \comrade\ ensures that \emph{(i)} a user cannot read a file from a shared repository unless that user
    has collected at least $t$ grants from $t$ distinct owners, and \emph{(ii)} a user cannot write a file
    in a shared repository unless that user is granted \wpermit{} access by $t$ of the file owners.

\item We build a prototype of \comrade{} and evaluate it within
    Amazon S3. We show that our solution scales well with the
    file size and the number of users.
\end{itemize}

The rest of the paper is organized as follows.
In Section~\ref{sec:model}, we introduce our notion of shared ownership in a file access control model.
In Section~\ref{sec:overview}, we detail our solution, \comrade{}, that enables the distributed enforcement of shared ownership in the cloud,
and we analyze its security.
In Section~\ref{sec:implementation}, we evaluate the performance of \comrade{} through an implementation within Amazon S3.
In Section~\ref{sec:disc}, we discuss further insights with respect to our proposals.
In Section~\ref{sec:related}, we review the related work, and we conclude the paper in Section~\ref{sec:conclusion}.

\section{SOM: Shared-Ownership File\\ Access Control Model}\label{sec:model}

In this section, we first formalize our notion of Shared Ownership within a
file access control Model named \dofac{}. We then discuss the main
shortcoming of centralized enforcement, and we define the
problem of \dofac{}'s distributed enforcement.

\subsection{Syntax and Semantics}

For simplicity, we do not consider directories (or other file
groupings) in \dofac. A file is created with the following request:
\begin{align*}
U\ \textbf{reqs}\ \mathrm{Create(\F, \textit{t}, \mathcal{O})}
\end{align*}

Upon receiving this request, \dofac{} creates a file $\F$,
assigns a user $U$ and all the users in $\mathcal{O}$ as $\F$'s
owners, and sets the file's threshold to $t$. \dofac{} grants
requests for file creation to authenticated users if the new
file name is unique.

To access a file, a user submits a request with an action he wishes to
perform on the file:
\begin{align*}
U\ \textbf{reqs}\ \mathrm{Action(\F)}
\end{align*}

\dofac{} does not instantiate concrete file actions, as these are left to the concrete
implementations, and we use $\mathrm{Action}(\F)$ to denote a generic file access
action on a file $\F$.

If an owner $O$ wishes to grant an action to $U$ over $\F$, then
he issues a credential of the form:
\begin{align*}
O\ \textbf{says}\ U\ \textbf{can}\ \mathrm{Action(\F)}
\end{align*}
Intuitively, a credential is a certificate by an owner to support an
action for a user. Figure~\ref{cgrammar} presents the full credential and request grammar.

File access requests are granted if and only if $t$
out of $n$ owners issue the corresponding
credentials. For example, if the threshold for $\F$ is 2, then $U$
can perform $\mathrm{Action}$ on $\F$ if the following credentials are
present:
\begin{align*} &O\ \textbf{says}\ U\ \textbf{can}\
\mathrm{Action(\F)}\\ &O'\ \textbf{says}\ U\ \textbf{can}\
\mathrm{Action(\F)} \end{align*} where $O$ and $O'$ are two of
$\F$'s owners.

The \dofac{} access control policy, $\Pi_\dofac$, is a mapping
from the set of all requests $\mathcal{R}$ and the set of
all credentials $\mathcal{C}$ into the $\{grant, deny\}$ decision set, such that the file threshold is always respected. Formally:

\begin{definition}[\textbf{Shared Ownership Access Control Policy}]
 The \dofac{} access control policy, denoted $\Pi_\dofac$, is a mapping $\mathcal{R} \times
\mathcal{C} \rightarrow \{grant, deny\}$ such that:
\begin{align*}
&\Pi_\dofac(U\ \textbf{reqs}\ \mathrm{Action}(\F), Creds) \mapsto grant\ \ \text{iff} \\
&\quad \{(O_1\ \textbf{says}\ U\ \textbf{can}\
\mathrm{Action}(\F)),\\
&\quad\ \ \dots,(O_t\ \textbf{says}\ U\ \textbf{can}\
\mathrm{Action}(\F))\} \subseteq Creds,\\
&\quad O_1 \neq \dots \neq O_t\ \text{and}\\
&\quad O_1 \in \mathcal{O}, \dots, O_t \in \mathcal{O},
\end{align*}
where $t$ is $\F$'s threshold, and $\mathcal{O}$ is the set
of $\F$'s owners.
\label{pol}
\end{definition}
We say that \dofac{} grants a file access to a given request $r$ and its
accompanying credentials $C$ if and only if $\Pi_\dofac(r, C) = grant$.

\begin{figure}[t]
\footnotesize
\begin{align*}
\textit{credential} ::=&
\ u\ \textbf{says}\ u\ \textbf{can}\ \mathrm{Action}(f)\\
\textit{request} ::=&\ u\ \textbf{reqs}\ (\mathrm{Create}(f, t, o)\ |\
\mathrm{Action}(f))\\
u::=&\ \mathrm{String}\\ t ::=&\ \mathbb{N}\\
f::=&\ \mathrm{String}\\
o::=&\ \mathrm{Set\ of\ Strings}
\end{align*}
\rule{\linewidth}{.7pt}
\caption{\textsf{SOM}'s credential and request grammar.\label{cgrammar}}
\end{figure}
Note that by Definition 1, \dofac{} treats
owners only as sources of credentials, and does not implicitly grant them any additional access rights.

\subsection{Centralized vs. Distributed Enforcement}

Existing credential-based access control systems have the following
enforcement model. A Policy Enforcement Point (PEP) has one designated
Policy Decision Point (PDP), which collects all the required
credentials and evaluates an access control policy (such as $\Pi_\dofac{}$) for a given request.
The PDP has one user who can administrate its access control policy.
We refer to this enforcement model as \emph{centralized} since
a single policy decision point grants all access requests.
Note that even if a PEP used multiple PDP components (managed by
different users), it would still require an additional PDP to
centrally decide how these decisions are combined.
In Appendix A, we show how \dofac{} can be centrally enforced. We do
this by specifying the \dofac{} access control policy as a Datalog
logic program. A Datalog interpreter can then act as the PDP.
Using Datalog interpreters as PDPs is also common in state-of-the-art access control systems (e.g., \cite{SecPAL}, \cite{DL}).

The key shortcoming of centralized enforcement is that the PDP's
access decision is solely sufficient for granting access requests.
The administrator, responsible for managing the rules
stored at the PDP, can therefore change the policy rules to grant requests that lack the necessary credentials.
In practice, this means that the shared ownership policy agreed upon by the owners can
always be bypassed, and thus the notion of shared ownership nullified.
The natural question to consider is how to enforce the \dofac{} policy
so that the agreed thresholds cannot be bypassed.
We refer to such an enforcement solution as \emph{distributed} because it must grant access
if and only if $t$ owners separately support the grant decision.

When considering a cloud as a collaboration platform,
this enforcement issue is even more difficult because a cloud platform
does not allow deployment of additional enforcement components.
A cloud platform only supports basic access control policies via
Access Control Lists (ACLs).
We frame these concerns as the
\dofac{} distributed enforcement problem.

\noindent \emph{Problem Statement:
How can the \dofac{} access control policy
be distributively enforced within a
cloud platform that supports only ACL-based PDPs?}
\newpage

We summarize the state-of-the-art with respect to this problem in
Table \ref{otable}.

\section{Commune: Distributed Enforcement of Shared Ownership}\label{sec:overview}
This section presents \comrade{}, a solution for distributed
enforcement of the \dofac{} access control policy.
As \dofac{} does not specify concrete file access operations,
we instantiate \comrade{} with \wpermit\ and \rpermit\ actions.
Before introducing our solution, we outline our cloud and attacker model.

\subsection{Cloud and Attacker Model}\label{}

We focus on a cloud storage platform, \server,
where a set of users $\mathcal{U}$ have personal accounts onto which they upload files.
We assume that each user authenticates himself to \server{} before downloading/uploading content.
A user $U\in\mathcal{U}$ can unilaterally decide who has access to files stored on his account.
In particular, \server{} allows each user to define access control policies of the type
$p: \mathcal{U} \times \{\wpermit,\rpermit\} \rightarrow \{grant, deny\}$.
We also assume that \server{} correctly enforces individual access control policies.
This model reflects the functionalities provided by existing cloud platforms, such as Amazon S3~\cite{AmazonS}.

Since we assume that \server\ authenticates users, we only focus on internal adversaries.
An adversary may try to gain \rpermit{} access to a file
even if fewer than $t$ owners have issued for her the corresponding credentials.
We refer to this adversary as a ``malicious reader''.
Alternatively, an adversary, who has been granted \wpermit{} access by
fewer than $t$ owners, may write a file $\F$ and then try to
publish $\F$ as if $\F$ were written by a writer who has been granted
\wpermit{} access by at least $t$ owners.
We refer to this adversary as a ``malicious writer''.
We also consider sets of users who collude to escalate their access rights.

We remark that, as is common for access control systems, we cannot prevent a user who
rightfully reads a file, from storing a copy to be read at a later time, even if his \rpermit\ rights have been revoked.

\subsection{Overview of \comrade}\label{subsec:overview}

We begin with the following observations:

\begin{observation}\label{one}
\emph{\comrade{}'s files cannot be stored on a single user account.}
\end{observation}
Following the discussion of centralized enforcement, no single
user account can be charged with making unilateral grant and deny decisions. Otherwise, a user may abuse his
rights and take unilateral access control decisions.
A straightforward solution where a file is encrypted (e.g., using a key shared among the owners) and the ciphertext is stored on a single account, allows that account holder to, for example, unilaterally deny \rpermit\ access to the ciphertext. If the ciphertext cannot be read, any mechanism to distribute or recover the encryption key is of no help.
 We argue, therefore, that \comrade{} cannot use a central repository because its administrator may unilaterally grant or deny access to the files stored therein.
Our alternative is to use a ``shared repository'', which is an abstraction built on top of the owners' personal accounts on \server.

\begin{table}[t]
\small
\centering
{\tabulinesep=.7mm
\begin{tabu}{l|c|>{\centering\arraybackslash}m{3.1cm}}
Enforcement/Ownership& \textbf{Individual} & \textbf{Shared} \\
\hline
\textbf{Centralized} & ACLs & Datalog-based Systems\\
\hline
\textbf{Distributed} & --- & \textsf{Commune} \\
\end{tabu}
}
\vspace{5pt}
\caption{Solutions for enforcing individual and shared ownership
within a cloud. Note that Datalog-based systems require modifications to the cloud platform.}
\label{otable}
\end{table}

\newpage
\begin{observation}\label{two}
\emph{\comrade{} cannot support in-place writing.}
\end{observation}
If \comrade{} were to allow in-place writing, then users who are granted \wpermit\ access could overwrite a file with ``garbage''.
This would equate to granting users the right to make unilateral decisions to delete the file, thus nullifying our efforts to prevent such scenarios.
A standard alternative to in-place writing is to introduce ``copy-on-write'' mechanisms whereby a new file is created upon each file \wpermit\ operation.
To optimize performance, \comrade{} implements versioning and splits files into \emph{units} (i.e., the unit of granularity of versioning) so that writing a new version of an existing file, only requires updating the units that have changed with respect to the previous version.

\begin{observation}\label{three}
\emph{\comrade{} cannot prevent users from disseminating a file through an out-of-band channel.}
\end{observation}
Access control solutions cannot prevent a user from distributing content through an out-of-band channel.
For example, a user who rightfully reads a file can leak it to third parties.
Similarly, a malicious writer can write a file and disseminate it through an out-of-band channel.
For example, a user can  publish files on his account on \server\ and make them available for others to read.
We cannot prevent such behaviour. \comrade{}, however, must at least allow honest readers, who abide to the
protocol specification, to distinguish between the content written by malicious writers and the content written by honest writers.

Given these observations, \comrade{} unfolds as follows.
At the system setup, users define the set of $n$ owners $\cal{O}$ and the threshold $t$.\footnote{Mechanisms to select the set of owners and the threshold $t$ are outside of our scope.
In settings like scientific collaboration scenarios, owners and
thresholds are agreed by the partners.}
\comrade{} abstracts the storage space of the owner's accounts on
\server{} as the ``shared repository''.  Each owner grants/denies
\rpermit{} and \wpermit{} access on his account to users
(including other owners) according to his individual access control policy.
The distributed enforcement of the \dofac{} access control policy then follows from
the enforcement of the individual access policies set by each owner.

To write a file to the shared repository, the writer encodes the file in \emph{tokens} and distributes the tokens to the owners' accounts.
A file is written to the shared repository if and only if the writer successfully distributes the file's tokens onto at least $t$ owners' accounts.
That is, a user has \wpermit\ access granted to the shared repository if and only if he has \wpermit\ access granted to at least $t$ of the owners' accounts.
We refer to such a user as an ``authorized writer''.

To read a file from the shared repository, the reader must fetch the file's tokens from at least $t$ distinct owners' accounts.
Therefore, a user has \rpermit\ access granted to the file if and only if he has \rpermit\ access granted to the file's tokens by at least $t$ owners.
We refer to such a user as an ``authorized reader''.

To secure access control enforcement,
\comrade{} is designed to fulfil the following properties.
\begin{itemize}
\item \textbf{Property 1: } A malicious writer (i.e., a user who has
not been granted \wpermit\ access to the shared repository by at least
$t$ owners), must not be able to write $\F$ and distribute it as if
$\F$ were authored by an authorized writer.
\item \textbf{Property 2: } A malicious reader (i.e., a user who has
not been granted \rpermit\ access to $\F$ by at least $t$ owners),
must not be able to recover the file contents.
    This property must also hold in case of \emph{revocation}.
    That is, the user must not be able to recover meaningful bits of the file even if he previously had \rpermit\ access to $\F$, and his access rights were later revoked.\footnote{Note that once a user has had access to a file, then he can locally store its copy. Similar to any other access control scheme, \textsf{Commune} cannot deter this behavior.} \comrade{} must also provide \emph{collusion resistance}. That is, coalitions of users---where no single user is an authorized reader---must not be able to pool their credentials to escalate their \rpermit\ access rights.
\end{itemize}

Property 1 ensures protection against malicious writers who try to disseminate content despite lacking the required credentials.
Property 2 guarantees that malicious readers cannot read content written to the shared repository.

\comrade{} fulfils Property 1 by design, through the abstraction of a shared repository and the copy-on-write mechanism (see Section~\ref{subsec:all}).
Property 2 is fulfilled through two cryptographic building blocks:
Secure File Dispersal (SFD), and Collusion Resistant Secret Sharing (CRSS).
SFD ensures that malicious readers cannot acquire any information about a file, even if they had previously access to the file and were later revoked.
CRSS builds atop SFD and ensures that coalitions of users where no single user has enough credentials to read the file, cannot pool their credentials in order to escalate their \rpermit\ access rights.

In the following, we describe and analyze SFD (Section~\ref{subsec:PAONT}) and CRSS (Section~\ref{subsec:CRSS});
in Section~\ref{subsec:all}, we detail the integration of both building blocks in \comrade.

\subsection{Secure File Dispersal (SFD)}\label{subsec:PAONT}

Information dispersal algorithms~\cite{Rabin} encode a file in $n$ chunks
so that any $t$ chunks (where $t\leq n$) are sufficient to decode it.
However, information dispersal algorithms do not provide any security
guarantees if the number of available chunks is less than $t$: any
party with less than $t$ chunks may still recover meaningful
information about the original file's content. This violates Property 2.

Previous work on securing information dispersal algorithms~\cite{resch11fast}
combines erasure codes with All-Or-Nothing Transformations (AONT)~\cite{rivest97fse}.
The latter is an efficient block-wise transformation that maps an $n$-block
bitstring in input to an $n'$-block bitstring in output (with $n'\geq n$).
AONTs are designed in such a way that, unless all the $n'$ output blocks are available,
it is hard to recover any of the input blocks.

Existing AONTs (e.g., \cite{rivest97fse,Boyko}) leverage block ciphers and rely on the secrecy
of a cryptographic key that is embedded within the output blocks.
Given all AONT output blocks, the key can be recovered; once the key is known, individual blocks can be reverted, independently of other blocks.
Therefore, current AONTs preserve their all-or-nothing property only for \emph{one time:}
knowledge of the key allows recovery of parts of the original data even if some of the output blocks are missing.
This is at odds with our security requirements.
As argued, we cannot prevent users from caching a local copy of the file and reading it at later time when their \rpermit\ rights may have been revoked.
However, we still want to provide revocation of a user who only stored the encryption key at the time when he had \rpermit\ access to the file.

We therefore introduce a new scheme, called Secure File Dispersal (SFD), that
combines information dispersal algorithms with an AONT scheme that
preserves its all-or-nothing property even if the adversary has the encryption key.

\noindent\textbf{Definition}\mbox{}

An SFD scheme consists of the following algorithms:

\begin{description}
    \item [$\{c_1,\ldots,c_n\} \leftarrow \mathrm{SFD.Encode}(t,n,\F,K,\lambda)$.]\mbox{}\\
    Encodes a file $F$ into $n$ chunks,
    such that $F$ can be correctly decoded using any $t$ chunks;
    $K$ denotes a key used in the encoding process and $\lambda$ is a security parameter.
    \item [$\F'\leftarrow \mathrm{SFD.Decode}(K,\mathcal{C},\lambda)$.]\mbox{}\\
    Takes as input a key $K$, a set of chunks $\mathcal{C}$, and security parameter $\lambda$; it outputs a file $\F'$.
\end{description}

\noindent \textbf{Correctness}\mbox{}

Given $\{c_1,\ldots,c_n\}\leftarrow\mathrm{SFD.Encode}(t,n,\F,K,\lambda)$,
we require that if $\mathcal{C}\subseteq\{c_1,\ldots,c_n\}$ and $|\mathcal{C}|\geq t$,
then\\ $\F\leftarrow \mathrm{SFD.Decode}(K,\mathcal{C},\lambda)$.

\noindent \textbf{Security}\mbox{}

We define the advantage of adversary $\cal A$ as follows:
\begin{align*}
&\mathrm{Adv}_{SFD}(\mathcal{A})=\mathrm{Pr}[\mathcal{A}(K,\mathcal{C})\!=\!f\ |\ K\!\gets\!\{0,1\}^{l}, l\!\geq\!\lambda, \F\!\equiv\!f_1,\!\ldots\\
&,\!f_m\!\gets\!\{0,1\}^{m\lambda},\{c_1,\!\ldots,\!c_n\}\!\gets\!\mathrm{SFD.Encode}(t,n,\F,K,\lambda),\\
&\mathcal{C}\subset\{c_1,\!\ldots,\!c_n\},|\mathcal{C}|\!<\! t,f\!\subseteq\! \F, |f|\!\geq\!\lambda].
\end{align*}
\noindent where $f\subseteq \F$ refers to a substring of $\F$.
We say that SFD is secure if, for any p.p.t. adversary, its advantage is negligible in the security parameter,
i.e., $\mathrm{Adv}_{SFD}(\mathcal{A})\leq negl(\lambda)$.

The above definition captures the scenario where the adversary is given access to enough chunks to decode $\F$,
and later her access rights are revoked.
Even though the adversary might have locally stored the secret key $K$, we require the probability
of $\cal A$ to recover any $\lambda$ consecutive bits of $\F$ to be
negligible in the security parameter, given less than $t$ output chunks.

\noindent\textbf{Instantiation}\mbox{}

Our SFD scheme combines information dispersal techniques with an all-or-nothing transformation inspired by Fast Fourier Transform.

\begin{figure}[t]
 \begin{center}
   \includegraphics[width=1.0\linewidth]{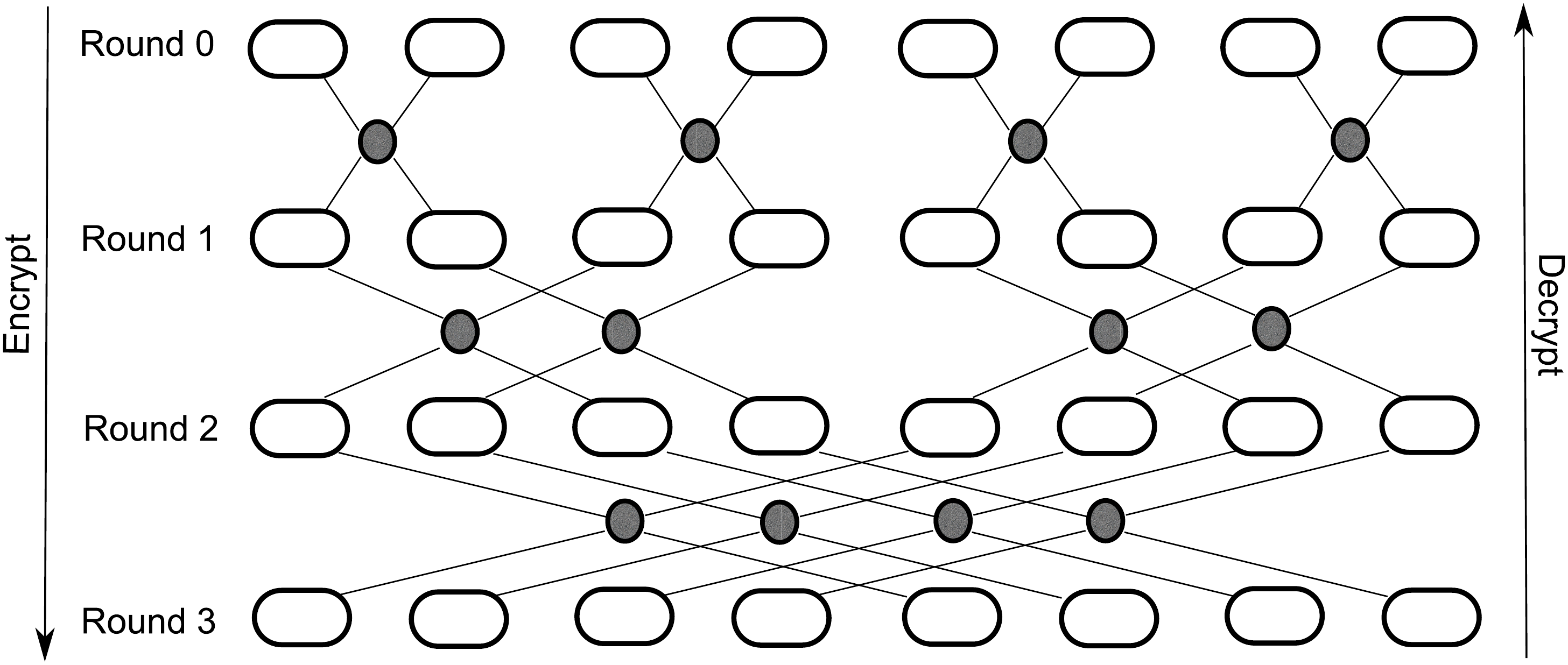}
 \caption{Sketch of the \aont{} scheme. We depict the case where the input of \aont{} consists of $m=8$ input blocks. Solid circles refer to the block cipher $E(\cdot)$, while empty circles depict its input/output blocks.\label{fig:paont}}
 \end{center}
\end{figure}

Let $E:\{0,1\}^{4\lambda} \rightarrow \{0,1\}^{2\lambda}$
be a semantically secure block cipher (e.g., $E(\cdot)$ could correspond to 256-bit $\mathrm{Rijndael}$~\cite{Rijndael}, with $\lambda=128$).\footnote{The key size is 2$\lambda$ and the input/output size is also $2\lambda$, totalling 4$\lambda$ size of input.}
The all-or-nothing transformation that SFD uses, and that we call \aont, takes as input a symmetric key $K$ (of size $2\lambda$) and $m$ input blocks $f_1,\ldots,f_m$ (each of size $\lambda$).
It executes in $\log_2{m}$ rounds and, at each round, applies $E(\cdot)$ to pairs of blocks. Each round is fed with the output of the previous round. Therefore, the original input $f_1,\ldots,f_m$ is treated as the output of round 0; similarly, the final output of the algorithm is the output of round $\log_2{m}$ (cf. Figure~\ref{fig:paont}).
The pseudo-code of \aont{} is shown in Algorithm~\ref{alg:paont}.
We omit the details of the decryption algorithm since it is specular to encryption.

Given the pseudo-code of \aont, our SFD scheme unfolds as follows:
\begin{description}
\item [$c_1,\ldots,c_n \leftarrow \mathrm{SFD.Encode}(t,n,\F,K,\lambda)$.]\mbox{}\\
    Parse $\F$ as $f_1,\ldots,f_m$  where each $f_i$ has size $\lambda$. Run $\bar{f}_1\ldots,\bar{f}_m\leftarrow\ant(K,f_1,\ldots,f_m)$.
    Use the information dispersal encoder to encode $\bar{f}_1\ldots,\bar{f}_m$ in $n$ chunks with
    reconstruction threshold $t$.\footnote{SFD can leverage any information dispersal algorithm (e.g., Reed-Solomon codes~\cite{lint82book}).}
\item [$\F'\leftarrow \mathrm{SFD.Decode}(K,\mathcal{C},\lambda)$.] \mbox{}\\
    Given $\mathcal{C}$, set of at least $t$ chunks, and key $K$, use the information dispersal decoder to decode blocks $\bar{f'}_1,\dots,\bar{f'}_m$.\\
    Run $f'_1\ldots,f'_m\leftarrow\ant(K,\bar{f'}_1,\ldots,\bar{f'}_m)$.
\end{description}

\noindent \textbf{Correctness}\mbox{}

If $\{c_1,\ldots,c_n\} \leftarrow\mathrm{SFD.Encode}(t,n,\F,K,\lambda)$, any
subset of at least $t$ chunks $\{c_{i_1},\ldots,c_{i_t}\}$ can be decoded into the whole output of \aont, namely $\bar{f}_1\ldots,\bar{f}_m$.
Given $K$, the output of \aont{} can be decrypted to recover $\F=f_1,\ldots,f_m$.

\noindent \textbf{Security}\mbox{}

Given the construction of our \aont{} scheme, it is easy to see that each input block depends on all output blocks and on the encryption key.
Furthermore, assuming that $E(\cdot)$ is a semantically secure block cipher, for any p.p.t.~algorithm $\cal A$, we have $\mathrm{Adv}_{SFD}({\cal A})\leq negl(\lambda)$. Refer to Appendix B for a security argument.

Note that a construct similar to \aont, was first mentioned by Rivest~\cite{rivest97fse} and later on used as a ``proof of storage'' in~\cite{vandijk12ccs}.
Nevertheless, ours is the first detailed FFT-inspired all-or-nothing transformation.
While \aont{} looks similar to~\cite{vandijk12ccs}, the construction proposed therein can use any pseudo-random permutation in the FFT network.
Our \aont{} requires a keyed-permutation, hence a block-cipher.
Furthermore, the goal of the adversary in~\cite{vandijk12ccs} is to recover, in a given amount of time, all \emph{output} blocks.
In contrast, the goal of our adversary is to recover any \emph{input} block even if he is not entitled to access the file.
This entails different security definition and analysis.

\begin{algorithm}[t]
  \caption{$\ant(K,f_1,\ldots,f_m)$  \label{alg:paont}}
  \begin{algorithmic}[1]
    \Statex
    \State Parse $f_1,\ldots,f_m$ as $f_1^0,\ldots,f_m^0$
    \For{$r \gets 1 \textrm{ to } \log_2{m}$} \Comment{round counter}
        \For{$i\gets 0 \textrm{ to } \frac{m}{2^r}-1$}
            \For{$j\gets 1 \textrm{ to } 2^{r-1}$}
                \State $f^r_{j+i\cdot 2^r} || f^r_{j+i\cdot 2^r+2^{r-1}}\gets E(K, f^{r-1}_{j+i\cdot 2^r}, f^{r-1}_{j+i\cdot 2^r+2^{r-1}})$
            \EndFor
        \EndFor
    \EndFor
    \State \Return{$f^r_1\ldots,f^r_m$ as $\bar{f}_1\ldots,\bar{f}_m$}
  \end{algorithmic}
\end{algorithm}

\subsection{Collusion Resistant Secret Sharing (CRSS)}\label{subsec:CRSS}

We now introduce our second building block, called Collusion Resistant Secret Sharing (CRSS).
Similar to secret-sharing, CRSS enables distributing a message among
a set of designated shareholders, so that any subset of shareholders of size equal to or greater than the threshold can reconstruct the secret.
CRSS, furthermore, allows users to collect ``delegations'' by shareholders to reconstruct the message.
If a user collects enough (i.e., above the threshold) delegations, he can rightfully reconstruct the message.
However, users cannot pool their delegations to reconstruct the message, unless one of them has collected enough delegations.
In \comrade, CRSS is used to secret-share the key $K$ used in SFD, in order to achieve collusion resistance.

CRSS is inspired by decentralized Attribute Based Encryption (ABE)~\cite{lewko11eurocrypt} where shares of a secret are \emph{blinded}
with shares of 0, such that, if a user collects enough shares for his identity, the blinding cancels out and the secret can be reconstructed.

\noindent \textbf{Definition}\mbox{}

Our definition of CRSS builds on top of a \emph{standard} threshold secret-sharing scheme $\mathrm{SS}$ with algorithms $\mathrm{SS.Share}$ and $\mathrm{SS.Combine}(\cdot)$, to share and reconstruct a secret, respectively.
We assume $\mathrm{SS}$ to be secure according to the \texttt{Game Priv} definition by Rogaway et al.~\cite{rogaway07ccs}.
That is, we assume that an adversary has only negligible advantage in identifying which out of two values was $(t,n)$ secret-shared using the $\mathrm{SS.Share}(\cdot)$ algorithm, even if the adversary can corrupt up to $t-1$ shareholders and access their shares.

CRSS defines the following algorithms:
\begin{description}
    \item[$\{s_1,\ldots,s_n\}\leftarrow \mathrm{CRSS.Share}(s,t,n)$.]\mbox{}\\
     Shares secret $s$ in a set of $n$ shares $\{s_1,\ldots,s_n\}$ with reconstruction threshold $t$.
    \item[$d_{i,j}\leftarrow \mathrm{CRSS.Delegate}(s_i,U_j)$.]\mbox{}\\
     Takes as input a share $s_i$ and an user identity $U_j$. The output is a \emph{delegation} $d_{i,j}$.
    \item[$s'\leftarrow \mathrm{CRSS.Combine}(\{d_{i_1,j},\ldots,d_{i_l,j}\}).$]\mbox{}\\
     Combines delegations $\{d_{i_1,j},\ldots,d_{i_l,j}\}$ into $s'$.
\end{description}

\noindent \textbf{Correctness}\mbox{}

Given $\{s_1,\ldots,s_n\}\leftarrow\mathrm{CRSS.Share}(s,t,n)$, then\\
$s\leftarrow\mathrm{CRSS.Combine}(\{d_{i_1,j},\ldots,d_{i_l,j}\})$
if\\ $d_{i_p,j}\leftarrow \mathrm{CRSS.Delegate}(s_{i_p},U_j)$ for $1\leq p\leq l$ and $l\geq t$.

\noindent \textbf{Security}\mbox{}

We model the security of CRSS using an adaptation of the \texttt{Game Priv} of~\cite{rogaway07ccs} and we denote the refined
game by $\texttt{Game Priv}^*$:

\begin{description}
\item \textbf{Init.} The adversary $\cal A$ submits two messages $x_0,x_1$ of equal length.
The challenger flips an unbiased coin $b$ and runs $\{s_1,\ldots,s_n\}\leftarrow \mathrm{CRSS.Share}(x_b,t,n)$.
\item \textbf{Find.} $\cal A$ can submit two types of queries. In Type-1 queries, the adversary can corrupt up to $t'\leq t-1$ shareholders and receives their shares.
At this time, $\cal A$ picks $t'$ indexes $i_1,\ldots,i_{t'}$ and receives $\{s_{i_1},\ldots,s_{i_{t'}}\}$.
In Type-2 queries, for any fresh identity $U_j$, the adversary can ask for up to $t''$ delegations, as long as $t'+t''\leq t-1$.
$\cal A$ submits an identity $U_j$ and $t''$ indexes $i_1,\ldots,i_{t''}$, and receives delegations $\{d_{i_1,j},\ldots,d_{i_{t''},j}\}$.
\item \textbf{Guess.} The adversary outputs her guess $b'$ and wins if $b'=b$.
\end{description}

We define the advantage of the adversary as the probability of its winning minus a half;
that is, $\mathrm{Adv}_{CRSS}^{\texttt{Priv}^{*}}({\cal A})=Prob[\texttt{Priv*}^{\cal A}]-\frac{1}{2}$. Therefore, we say that CRSS is secure if any p.p.t.~algorithm ${\cal A}$ has only negligible advantage in winning \texttt{Game Priv*}.

The above \texttt{Game Priv*} models a scenario where a set of malicious users, including up to $t'$ shareholders, collect up to $t''$ delegations for each of their identities.
If $t'+t''\geq t$, the malicious shareholders can produce the missing delegations for any of the colluding user identities,
so that the secret can be reconstructed by means of $\mathrm{CRSS.Combine}(\cdot)$.
Otherwise, colluding users must not be able to retrieve the secret.

\noindent \textbf{Instantiation}\mbox{}

Our CRSS scheme is based on the threshold secret-sharing scheme proposed in~\cite{charnes94ccs}, which is defined as follows:

\begin{description}
\item[$g^x, \{x_1,\ldots,x_n\}\leftarrow \mathrm{SS.Share}(-,t,n)$.]\mbox{}\\
    Pick a cyclic group $G$ of prime order $q$ where the discrete logarithm assumption holds; let $\langle g\rangle=G$.
    Pick a random $x\in Z_q$ and set the secret to $g^x$.
    Pick a random $t-1$-degree polynomial $X$ with coefficients in $Z_q$, such that $X(0)=x$. Set the $i$-th share to $x_i=X(i)$.
\item[$s'\leftarrow \mathrm{SS.Combine}(\{x_{i_1},\ldots,x_{i_l}\})$.]\mbox{}\\
    Given shares $\{x_{i_1},\ldots,x_{i_l}\}$, use polynomial interpolation to
    recover the secret. That is:
        \[
        s'=g^{\sum_{p=1}^{p=l} x_{i_p}\lambda_{p}}\nonumber
        \]
    where $\lambda_{p}=\prod_{1\leq k\leq l}^{k\neq p}\frac{x_{i_k}}{x_{i_k}-x_{i_p}}$.
\end{description}

Note that in the above scheme, the secret is not given as input to the $\mathrm{Share}$ algorithm; rather, it is set to $g^x$ for a randomly chosen $x$.
Given the above algorithms, our CRSS scheme unfolds as follows:

\begin{description}
\item[$\{s,s_1,\ldots,s_n\}\leftarrow \mathrm{CRSS.Share}(-,t,n)$.] \mbox{}\\
    Run $g^x,\{x_1,\ldots,x_n\}\leftarrow \mathrm{SS.Share}(-,t,n)$.
    Pick $H(\cdot):\{0,1\}^*\rightarrow G$ to be a cryptographic hash function that maps random strings in $G$.
    Pick a random $t-1$-degree polynomial $Y$ with coefficients in $Z_q$, such that $Y(0)=0$, and denote $y_i=Y(i)$.
    The secret is set to $s=g^x$ while each share is set to $s_i=(x_i,y_i)$.
\item[$d_{i,j}\leftarrow \mathrm{CRSS.Delegate}(s_i,U_j)$.] \mbox{}\\
    Parse $s_i=(x_i,y_i)$ and output $d_{i,j}=g^{x_i}H(U_j)^{y_i}$.
\item[$s'\leftarrow \mathrm{CRSS.Combine}(\{d_{i_1,j_1},\ldots,d_{i_l,j_l}\})$.]\mbox{}\\
     Run $s'\leftarrow \mathrm{SS.Combine}(\{d_{i_1,j_1},\ldots,d_{i_l,j_l}\})$ and output $s'$.
\end{description}

\noindent \textbf{Correctness}\mbox{}

If $l\geq t$, then $\mathrm{CRSS.Combine}(\{d_{i_1,j},\ldots,d_{i_l,j}\})$ outputs
\begin{eqnarray}
        s'&=&\prod_{p=1}^{p=l} \left(d_{i_p,j_p}\right)^{\lambda_{i_p}}\nonumber\\
        &=&\prod_{p=1}^{p=l} \left(g^{x_i}H(U_j)^{y_i}\right)^{\lambda_{i_p}}\nonumber\\
        &=&g^{\sum_{p=1}^{p=l}\lambda_{i_p} x_{i_p}}H(U_j)^{\sum_{p=1}^{p=l}\lambda_{i_p} y_{i_p}}\nonumber\\
        &=&g^k H(U_j)^0\nonumber\\
        &=&g^k\nonumber\\
        &=&s.\nonumber
\end{eqnarray}

\noindent \textbf{Security}\mbox{}

The security of CRSS is based on the fact that, in the random oracle model,
delegations for different identities cannot be combined to remove the blinding factor from the secret.
Assuming that $H(\cdot)$ is modeled as a random oracle and that the discrete logarithm assumption holds in $G$,
we can show that any p.p.t.~algorithm $\cal A$ has only negligible advantage in winning \texttt{Game Priv*}.
We provide details of the proof in Appendix C.

\subsection{Commune: Protocol Specifications}\label{subsec:all}

Recall that \comrade{} leverages a shared repository, which is an abstraction of the owners' storage space on \server.
The shared repository uses a versioning system so that content cannot be overwritten but only new content can be added.
In particular,  \comrade{} optimizes performance by splitting a file in smaller \emph{units}, and encoding/decoding each unit separately.
Therefore, when a new file version is written to the shared repository, the writer only needs to upload the units that have changed from the previous version.

Files written to the repository are encoded in \emph{tokens} and distributed across the owners' accounts.
Leveraging the basic ACLs of \server, owners exert their individual policy on the tokens they store on their accounts.
The distributed enforcement of the \dofac{} policy is implied by the enforcement of each owner's individual policy on his tokens by \server.

Encoding must guarantee both correctness and security of reading operations.
Hence, users who are authorized to read at least $t$ tokens must be able to decode the original file; users who are granted \rpermit\ access
on less than $t$ tokens must not be able to recover its content.
Furthermore, users must not be able to pool their credentials to
escalate their access rights.

\noindent\textbf{Create a File}\mbox{}

File creation requires one user, the file creator, to ``bootstrap'' the system and write the initial version of the file into the repository.
For this reason, we assume that---at the file creation time---the file creator has been granted the right to write new data to each of the owner's accounts on \server.

The file creator first divides the file $\F$ into $k$ fixed-sized units.
For each unit $F_i$ $(i \in [1,\ldots,k])$, he runs
$\{s_i,s_{i1},\ldots,s_{in}\}\leftarrow \mathrm{CRSS.Share}(-,t,n)$ to produce a fresh secret $s_i$ and $n$ of its shares.
Secret $s_i$ is used as a symmetric key to encode the unit $\F_i$ in $n$ chunks using SFD.
That is, the file creator runs $\{c_{i1},\ldots,c_{in}\} \leftarrow\mathrm{SFD.Encode}(t,n,\F_i,s_i,\lambda)$.
The token of the unit $\F_i$ for the owner $O_j$ is set to $(c_{ij},s_{ij})$ (i.e., one chunk outputted by $\mathrm{SFD.Encode}(\cdot)$ and one secret-share outputted by $\mathrm{CRSS.Share}(\cdot)$).
Finally, for each owner $O_j$, the file creator writes
$\{(c_{ij},s_{ij})\}_{i \in [1,\ldots,k]}$ to $O_j$'s account on \server.
Each owner, therefore, receives one token for each unit that constitutes $\F$.

\noindent\textbf{Grant/Deny Write Rights}\mbox{}

An owner $O_j$ grants \wpermit\ rights to a user $U_l$ by granting to $U_l$ the right to write new data (i.e., tokens) to his account.
Similarly, $O_j$ denies \wpermit\ rights to $U_l$ by denying $U_l$ the right to write new data to his account.

\noindent\textbf{Update a File}\mbox{}

Assume $U_l$ wants to write a new version of a file $\F$.
For simplicity, assume that the new version differs from the previous one by only one unit $F_i$
(the case where the old and the new versions differ in several units is handled in a similar fashion).

At this point, some owners may allow $U_l$ to write tokens to their accounts while others may not.
Let $\cal{O}^+$ be the subset of owners who grant to $U_l$ \wpermit\ rights to their accounts.
Similarly, let $\cal{O}^-$ be the subset of owners who deny to $U_l$ \wpermit\ rights to their accounts.
$U_l$ can, therefore, only distribute tokens to owners in $\cal{O}^+$.
This scenario is equivalent to the case where $U_l$ distributes tokens to all owners in $\cal{O}$,
but the ones in $\cal{O}^-$ decide to reject the version produced by $U_l$ and make the received tokens unavailable.

$U_l$ is an authorized writer and his version accepted (i.e., considered as written to the shared repository) if and only if $|\mathcal{O}^+|\geq t$.
In this case, there are at least $t$ tokens for the new unit, so it may be decoded by users who collect enough credentials.
If $|\mathcal{O}^+|< t$, user $U_l$ is not authorized to write and his version is rejected (i.e., considered as not written to the repository),
since there are not enough tokens to successfully decode the unit produced by $U_l$.
\newpage

\noindent\textbf{Grant/Deny Read Rights}\mbox{}

Recall that for each unit $\F_i$, an owner $O_j$ receives the token $(c_{ij},s_{ij})$.
$O_j$ can grant to $U_l$ \rpermit\ access to that unit by \emph{endorsing} the token for $U_l$ and granting to $U_l$ \rpermit\ access on the endorsed token.
Token endorsement requires $O_j$ to run $d_{ij,l}\leftarrow\mathrm{CRSS.Delegate}(s_{ij},U_l)$.
The endorsed token $(c_{ij},d_{ij,l})$ is then made available by $O_j$ for $U_l$ to read.
If a file consists of multiple units, $O_j$ must endorse all relative tokens for $U_l$ and grant to $U_l$ \rpermit\ access on all endorsed tokens.

$O_j$ can revoke \rpermit\ rights that were previously granted, by denying to $U_l$ the right to read the previously endorsed tokens.

\noindent\textbf{File Reading}\mbox{}

\begin{figure}[tb]
\begin{center}
        \includegraphics*[width=0.8\linewidth]{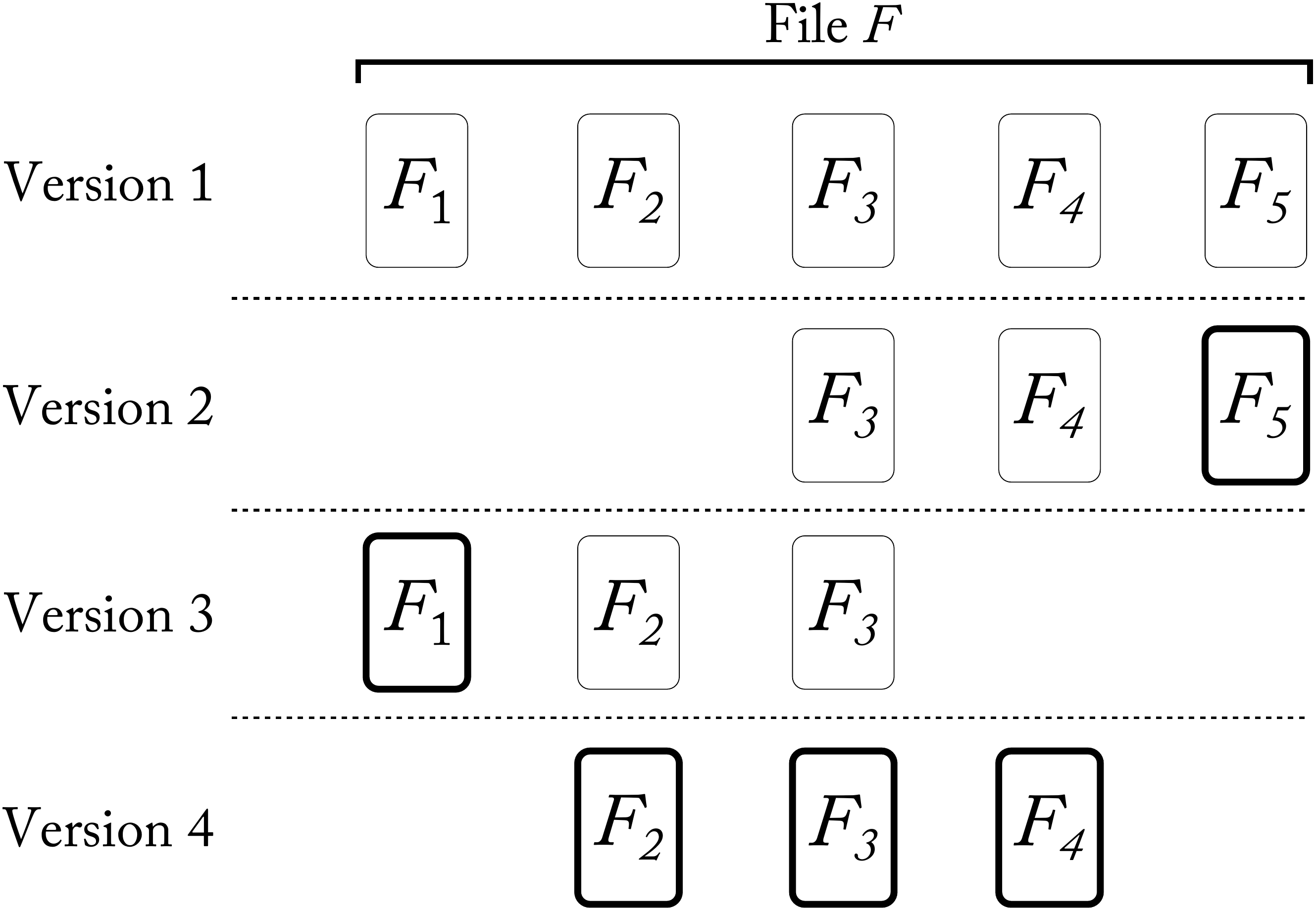}
\end{center}
\caption{Reading a file in \textsf{Commune}. To read version $x$ of a file, a reader must fetch the latest endorsed tokens created up to (and including) version $x$, for each unit that comprises the file. The figure shows a file that spans 5 units (depicted as boxes) and has 4 versions.
Reading version 4 requires fetching (a) tokens that encode units that are new in this version, (b) tokens that encode $\F_1$ created at version 3, and (c) tokens that encode $\F_5$ created at version 2. All of these tokens are shown with tick borders.
}
\label{fig:version}
\end{figure}

If the original file spans several units, $U_l$ must decode each unit separately in order to read the entire file.
That is, for each unit, he uses the set of endorsed tokens fetched to recover the secret key via $\mathrm{CRSS.Combine}(\cdot)$ and then uses the secret key to decode the unit via $\mathrm{SFD.Decode}(\cdot)$.
Note that for an authorized reader to read version $x$ of file $\F$, he must fetch the latest endorsed tokens created up to (and including) version $x$, for each unit that comprises the file. This is illustrated in Figure~\ref{fig:version}.

Assume the user $U_l$ is granted read access to $\{ (c_{{ij}_1},d_{{ij}_1,l}),\ldots, (c_{{ij}_t},d_{{ij}_t,l})\}$.
To access unit $\F_i$, he runs  $s_i\leftarrow \mathrm{CRSS.Combine}(\{d_{{ij}_1,l},\ldots,d_{{ij}_t,l}\})$
to recover secret key $s_i$ and then runs $\F_i\leftarrow \mathrm{SFD.Decode}(s_i,\{c_{{ij}_1},\ldots,c_{{ij}_t}\},\lambda)$.
$U_j$ proceeds in a similar way in order to recover all units of $\F$ that he has access to.

\noindent \textbf{Security analysis}\mbox{}

From Sections~\ref{subsec:PAONT} and~\ref{subsec:CRSS}, it follows that given $t$ tokens of a file unit $\F_i$, endorsed  for a unique user identity,
it is possible to recover both the secret key used to encode $\F_i$ and its \aont{} ciphertext, so that the original file can be decrypted.
That is, honest readers can access files written by honest writers, if they are granted such right by at least $t$ out of $n$ owners.

Property 1 (cf. Section~\ref{subsec:overview}) is fulfilled as follows.
First, \comrade{} uses copy-on-write to prevent writers from overwriting content in the shared repository with garbage.
Second, malicious writers (i.e., writers who have been
granted \wpermit\ access by fewer than $t$ owners) are unable to distribute a file without honest readers detecting it.
In other words, a file is considered as written if and only if it is correctly encoded in tokens and those tokens are distributed to and endorsed by at
least $t$ out of $n$ owners. Any content distributed through other
means (out of band channels, the writer's account on \server, etc.) is
easily recognized as malicious by honest readers.
We argue that the detection is the only solution for protecting honest
readers, because there are no mechanisms to deter malicious writers from disseminating any content (cf. Observation 3).
We also stress that honest readers can easily detect writers that distribute polluted (i.e., non-decodable) tokens.
Denial-of-service attacks, therefore, are out of our scope.

Property 2 is satisfied by combining CRSS and SFD.
The former ensures that coalitions of users, where no single user has enough tokens endorsed for his identity,
pool their endorsed tokens in order to escalate their access rights.
The latter addresses the case where at a time $\tau_1$ a user has
access to $t$ or more endorsed tokens of a file unit $\F_i$, but at a
time $\tau_2>\tau_1$, his access rights are revoked. That is, at the
time $\tau_2$, the user has access to fewer than $t$ endorsed tokens.
SFD  ensures that even if, at the time $\tau_1$ the user may have
cached the key used to encode $\F_i$ in tokens, he will not be able
at the time $\tau_2$ to decode parts of $\F_i$.
Note that, once a user has access to the file, then he can locally store any plaintext content of his choice.
Similar to other access control schemes, \comrade{} cannot deter this behavior.

Finally, given the guarantees that \comrade{} makes for enforcing the
\wpermit{} and \rpermit{} actions, it follows that \comrade{} is a
(correct) solution for distributed enforcement of the \dofac{} access
control policy (see Definition \ref{pol}).
%


\section{Prototype Design \& Evaluation}\label{sec:implementation}

To evaluate \comrade{}, we implement a prototype integrated with Amazon S3~\cite{AmazonS}.
In this section, we describe the implementation and evaluate its performance.

\subsection{Implementation Setup}

We leverage Amazon S3 to instantiate \server: for each user in $\cal{U}$, we create personal accounts in Amazon S3, into which users
can upload content and define arbitrary access control policies.
In our implementation, we use Amazon S3 access control features to distribute tokens from the file creator to the set of owners $\cal{O}\subseteq\cal{U}$. In particular, we assume that each user sets up \emph{(i)} one ``temporary'' folder where other peers are granted $\mathrm{write}$ access, and \emph{(ii)} one ``main'' folder where endorsed tokens are stored and retrieved. When the file creator wants to distribute a token to owner $O_j$, he writes the token to $O_j$'s temporary folder.
Since no other user apart from $O_j$ has $\mathrm{read}$ access to the temporary folder, the new token is protected from unauthorized access.
At this point, $O_j$ can endorse the token for any other user, say $U_l$, store the endorsed token in his main folder and grant $\mathrm{read}$ access rights on the endorsed token to $U_l$.

Our prototype, implemented in Java, is a multi-threaded client-side interface to repositories hosted on Amazon S3.
The client runs on a user's machine and uploads/down\-loads content to/from the repositories.
For evaluation, we use Ubuntu \emph{v. 12.04} on a machine running Intel Core i5-2400 (at 3.10 GHz),
where up to 4 GB of RAM are allocated to the OpenJDK VM.

The implementation of SFD leverages Rijndael~\cite{Rijndael} (implemented using the Bouncy Castle Java library~\cite{Bouncy})
as the underlying block cipher for \aont{} and systematic Reed-Solo\-mon codes~\cite{lint82book} (implemented using the Jerasure library~\cite{Jerasure1,Jerasure2})
for information dispersal. We chose a symbol size of $16$ bytes, and a security parameter $\lambda=128$ bits.
Our implementation of CRSS leverages the $\mathrm{secretsharejava}$ library~\cite{secretsharejava} with a 386-bit modulus.

To optimize performance, our prototype handles file unit operations at a smaller granularity, called \emph{pieces}.
During the creation of any file unit, the unit is split into pieces that are processed in parallel.
A token for each unit contains one output chunk of SFD for each piece that composes the unit.
The piece size $w$ is chosen such that $t \lambda | w$, where $\lambda$ is the security parameter and $t$ is the required reconstruction threshold.
This condition ensures that
\emph{(i)} a piece can be encrypted in an integer number of ciphertext blocks of $\lambda$ bits,
\emph{(ii)} an encrypted piece can be divided into an integer number of input chunks for the Reed-Solomon encoder,
and \emph{(iii)} the size of each chunk of the Reed-Solomon encoder/decoder is at least $\lambda$ bits.

\begin{figure*}[tb]
\begin{center}
    \subfigure[Impact of the piece size.]{
        \label{fig:chunk}
        \includegraphics*[width=0.47\linewidth]{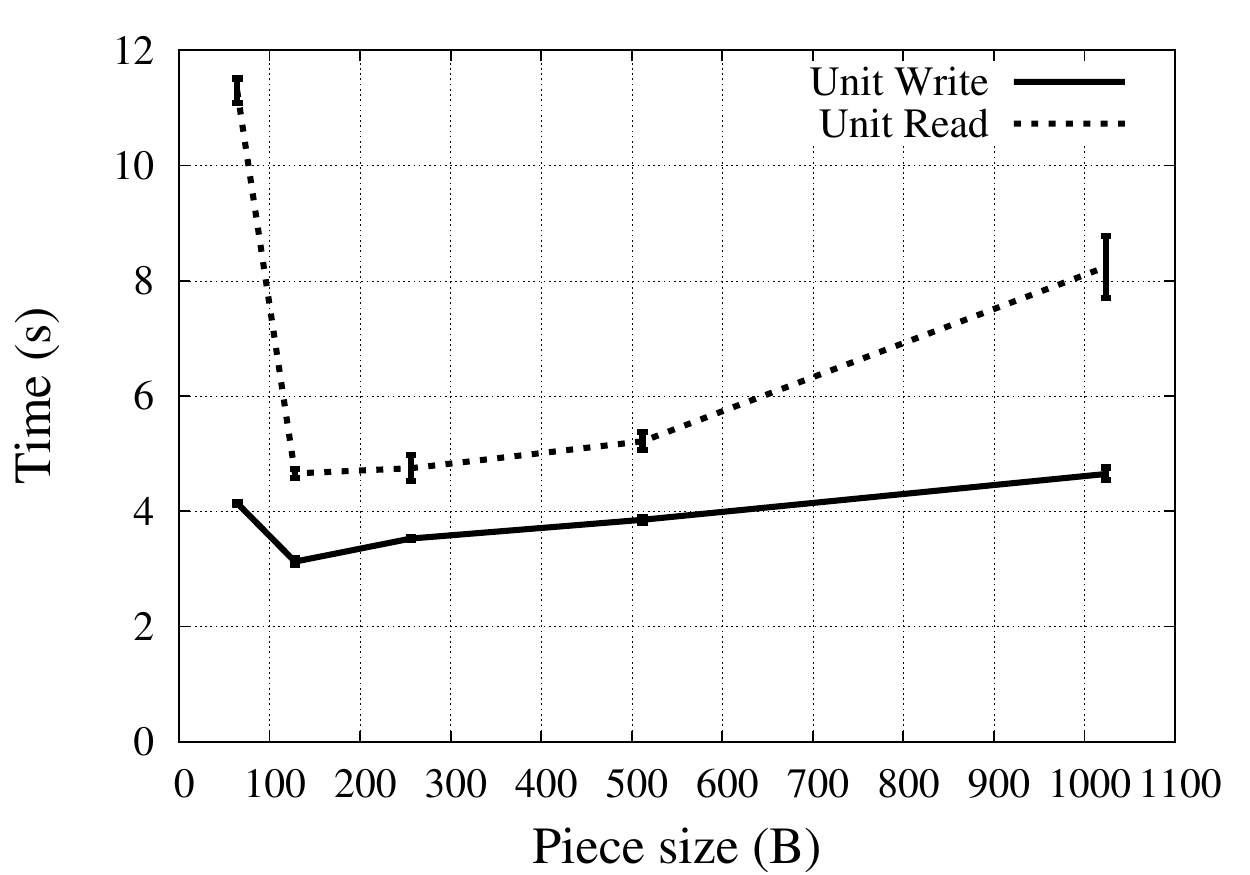}
    }
     \subfigure[Impact of the reconstruction threshold.]{
        \label{fig:threshold}
        \includegraphics*[width=0.47\linewidth]{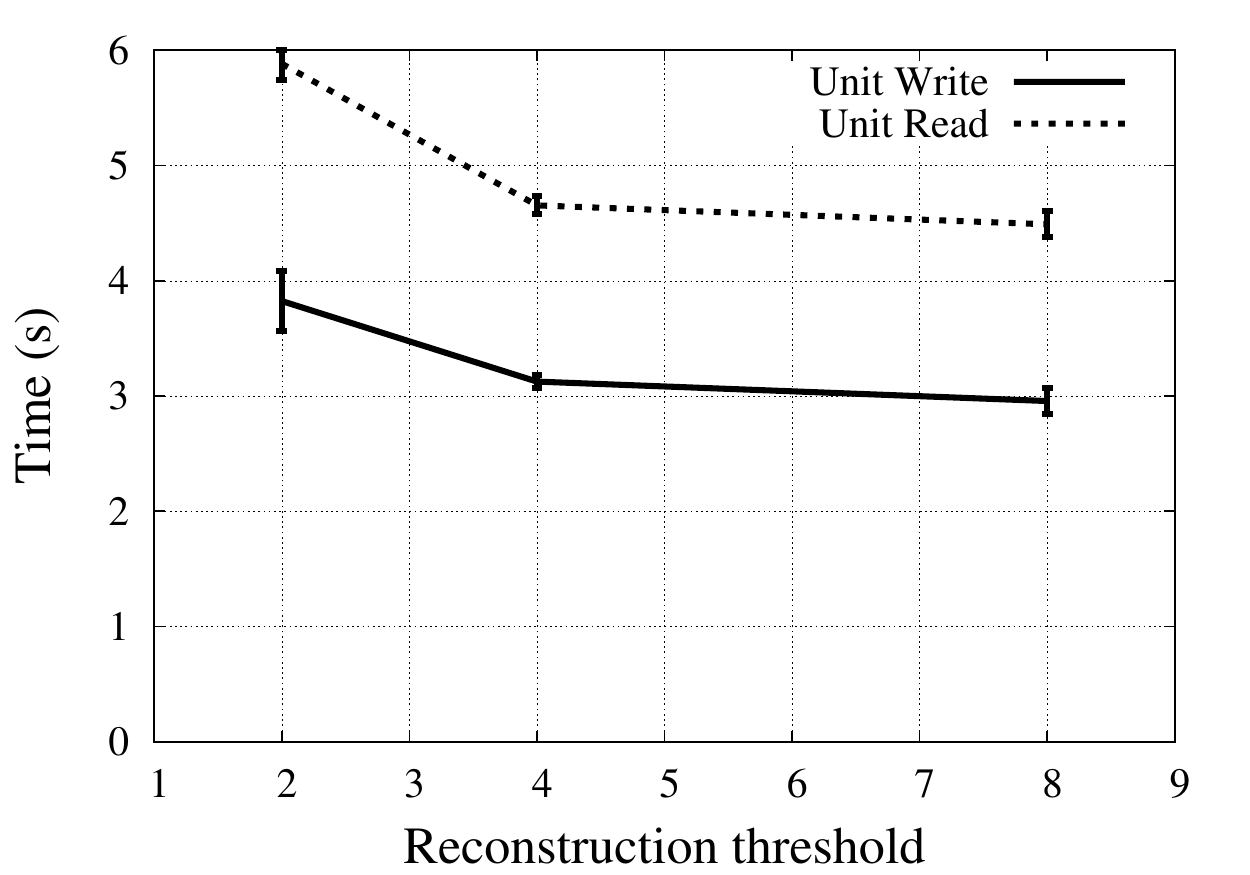}
    }
      \subfigure[Impact of the number of owners.]{
        \label{fig:users}
        \includegraphics*[width=0.47\linewidth]{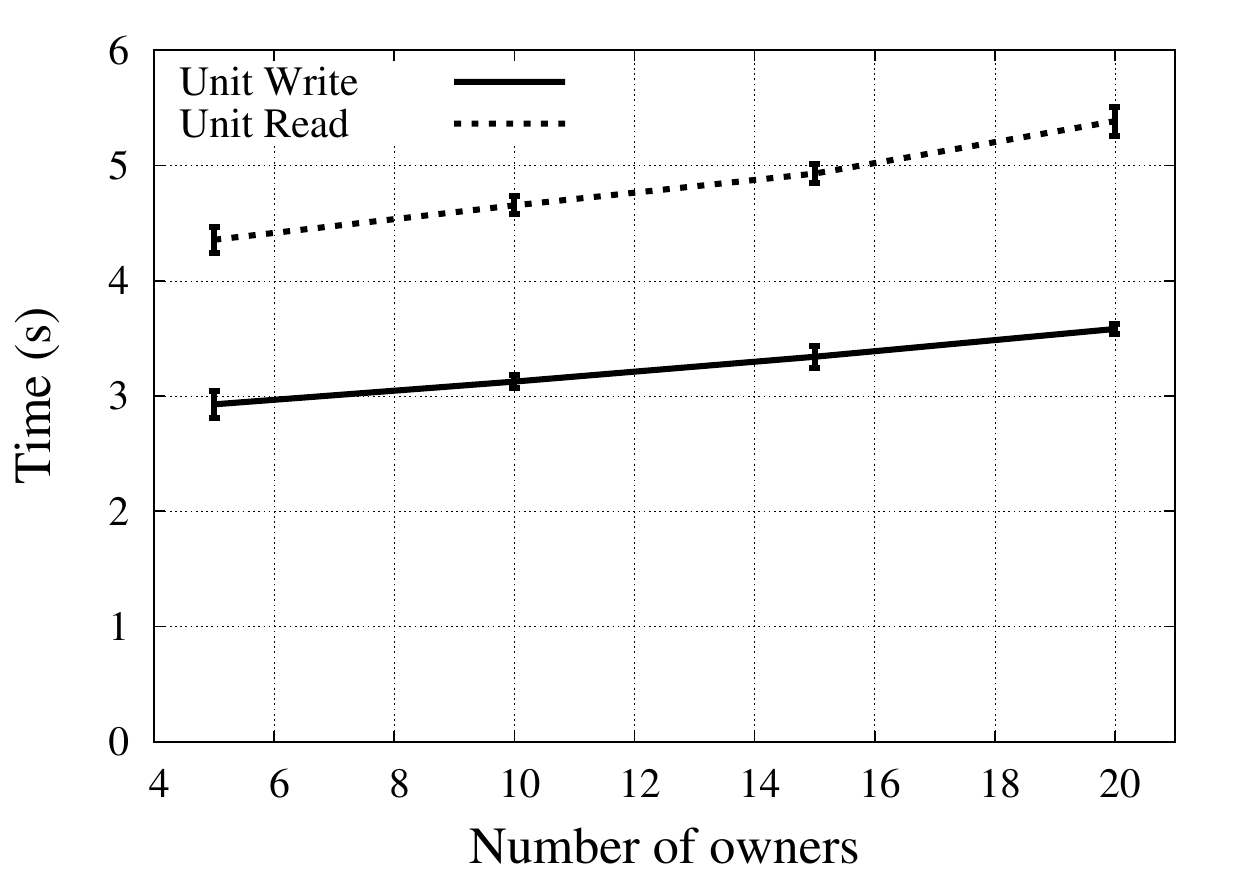}
    }
        \subfigure[Impact of the unit size.]{
        \label{fig:file}
        \includegraphics*[width=0.47\linewidth]{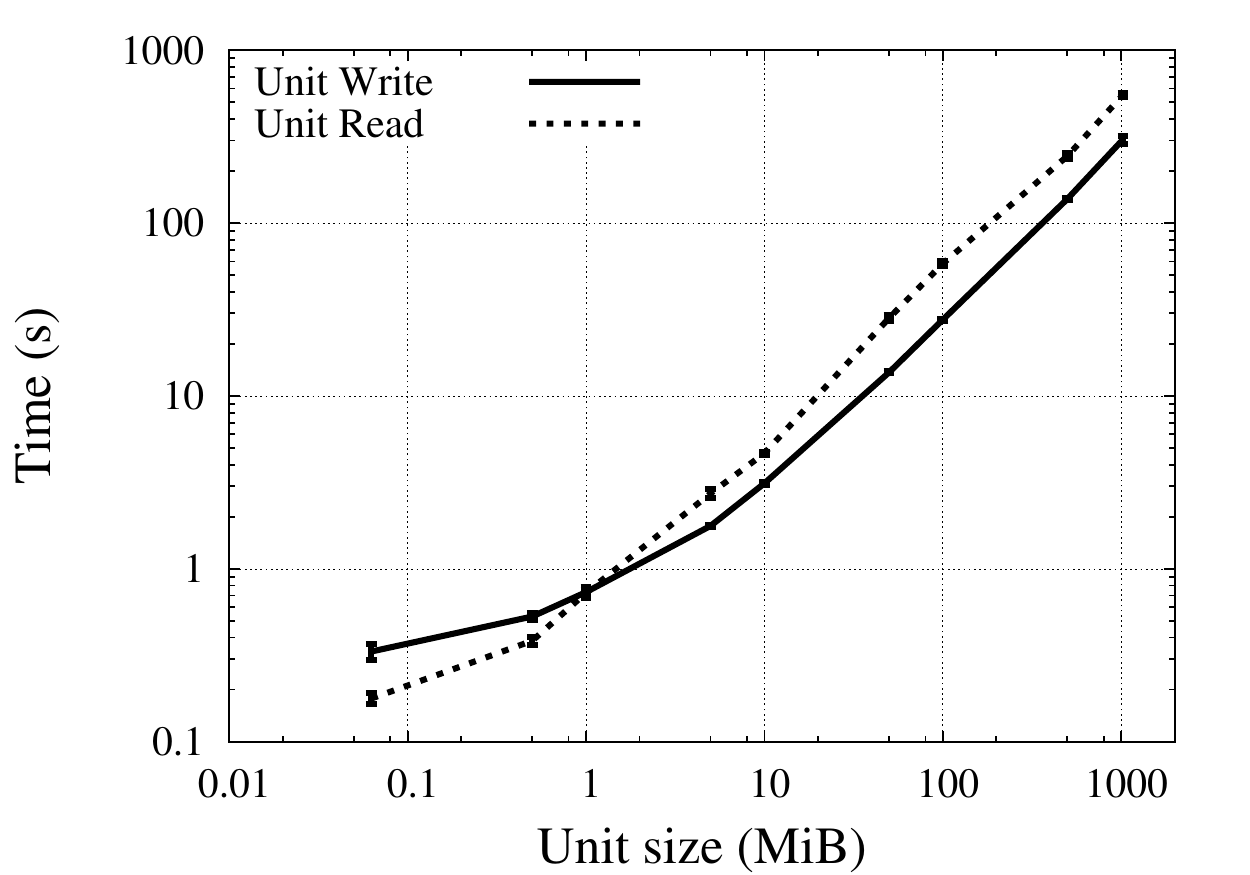}
    }
\end{center}
 \caption{Performance evaluation of our prototype implementation. The system parameters were chosen from Table~\ref{tab:nominal}. Each data point in these plots is averaged over 20 independent measurements; where appropriate, we also provide the corresponding 95\% confidence intervals.}
 \label{fig:results}
 \end{figure*}

\begin{figure*}[tb]
\begin{center}
        \includegraphics*[width=\linewidth]{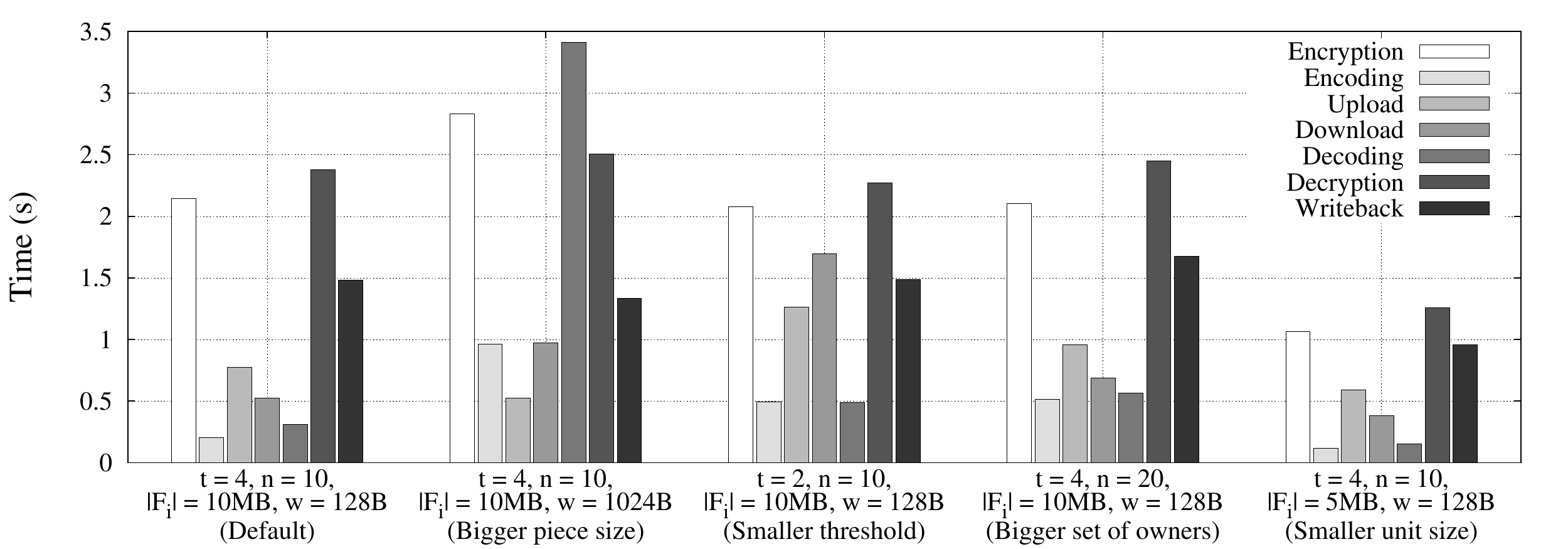}
    \caption{Runtime analysis for five different configurations of our prototype
		implementation. The label ``writeback'' denotes the operation of writing the data back on disk onto the file descriptor (during Unit Read). Each data point is averaged over 20 different measurements.}
    \label{fig:histograms}
\end{center}
\end{figure*}

\subsection{Evaluating Single Unit Write/Read}\label{sec:evaluation}

\begin{table}
\centering
\scalebox{1.5}{
\begin{tabular}{|c|c|}
  \hline
  \textbf{Parameter} & \textbf{Default Value}\\
  \hline
  \hline
  $w$           & 128 B\\
  $t$           & 4 \\
  $n$           & 10 \\
  $|F_i|$         & 10 MiB\\
  \hline
\end{tabular}
}
\vspace{5pt}
\caption{Default parameter values used throughout the evaluation.}
\label{tab:nominal}
\end{table}

We evaluate the performance of \comrade{} for a single file unit write and read, with respect to
\emph{(i)} the piece size $w$,
\emph{(ii)} the reconstruction threshold $t$,
\emph{(iii)} the number of owners $n$, and
\emph{(iv)} the size of a file unit $|F_i|$.
For that purpose, we assume the default parameter values shown in Table~\ref{tab:nominal}.\footnote{Since our SFD scheme requires that the number of plaintext blocks input to \aont{} is a power of 2, we also set $w$, $\lambda$, and $t$ to be powers of 2, in order to ensure that $t\lambda|w$.}
We then change one variable at a time, to assess its impact on the system performance.
For each configuration, we measure the time required \emph{(i)} to create and upload a unit $F_i$, denoted by \emph{Unit Write}
in our plots, and \emph{(ii)} to retrieve $F_i$, denoted by \emph{Unit Read}.
These times are measured from the initiation of the operation until the output is
available either in the repositories (for unit write) or on a local disk (for unit read).
During the experiments, we controlled for the effect of caching by uploading random binary streams at each run.

During \emph{Unit Read}, our client fetches endorsed tokens from $t$ randomly chosen owners (who granted \rpermit\ access right on their endorsed tokens).
Recall that a $(t,n)$ systematic erasure code outputs $t$ data chunks and $n-t$ parity chunks. Since data chunks need not be processed at decoding time, our implementation therefore accounts for the average-case scenario where the probability that a retrieved token contains a data chunk is bounded by $t/ n$.

Note that we do not evaluate the time required to grant \rpermit\ rights (i.e., the time required to endorse a token), since it does not depend on any of the considered parameters.

Our results are depicted in Figure~\ref{fig:results}. Each data point
in our plots is averaged over 20 runs; where appropriate, we also
provide the corresponding 95\% confidence intervals. For better analysis, we also monitor the runtime incurred in intermediate steps for a number of configurations; our results are illustrated in Figure~\ref{fig:histograms}.

Our findings show that writing a new unit (\emph{Unit Write}) is less expensive than reading it (\emph{Unit Read}).
We speculate that the bottleneck for \rpermit\ operations is due to the unique file descriptor that multiple parallel threads share when writing decoded pieces to local disk.

\noindent \textbf{Impact of the Piece Size}\mbox{}

Figure~\ref{fig:chunk} shows the impact of the piece size $w$ on the  performance of \comrade{}.
Smaller $w$ leads to a smaller number of input blocks to the \aont{} scheme, which results in better performance.
Recall that \aont{} requires $\log_2{m}$ rounds of encryption over all the $m$ input blocks.
However, we experience higher latencies for very small values of $w$, especially in the \emph{Unit Read} operation.
This is due to the overhead incurred by different threads that must synchronize on the single file descriptor in order to write data to disk.

Throughout the rest of the evaluation, we set $w=128$ bytes, since it shows to be a good trade-off between \aont{} performance and disk latency.

\noindent \textbf{Impact of the Reconstruction Threshold}\mbox{}

Figure~\ref{fig:threshold} shows the impact of the reconstruction
threshold $t$ on the system's performance.
As $t$ decreases, the chunk size of the Reed-Solomon encoder increases;
this results in larger chunk upload and download times.
Figure~\ref{fig:histograms} also shows that a smaller value of $t$ results in longer encoding and decoding times.
On the one hand, during \emph{Unit Write}, small values of $t$ result in larger encoding overhead since the size of the encoding matrix considerably increases.
On the other hand, during \emph{Unit Read}, small values of $t$ decrease the probability of recovering data chunks (w.r.t. the probability of recovering parity chunks),
which makes decoding slower (cf. Figure~\ref{fig:histograms}).

\noindent \textbf{Impact of the Number of Owners}\mbox{}

Figure~\ref{fig:users} shows that latency increases for both \emph{Unit Write} and \emph{Unit Read} as the number of owners grows.
During \emph{Unit Write}, this increase is because the file creator must distribute tokens to more peers.
Latency increase during \emph{Unit Read} is due to the overhead of the Reed-Solomon decoder, which must decode parity chunks.
As mentioned earlier, the Reed-Solomon encoder outputs $t$ data chunks and $n-t$ parity chunks.
Data chunks do not need decoding while parity chunks need to be decoded and increase the decoder overhead.
As $n$ grows, the number of parity chunks increases, and so does the probability of fetching one of them.
As the number of fetched parity chunks increases, decoding becomes slower.

\noindent \textbf{Impact of the Unit Size}\mbox{}

Figure~\ref{fig:file} shows \comrade{}'s performance for different unit sizes.
In particular, we vary the number of pieces, that comprise a file unit.
Our results show that the time required to write/read units in \comrade{} increases almost linearly with the unit size.
(Note that Figure~\ref{fig:file} uses semi-logarithmic axes.)
The time required to read a 10 MB unit in our implementation is roughly 4.47 seconds.
In this case, the effective throughput of our prototype is close to 18 Mbps.

To optimize the performance, the choice of the unit size should depend on the user update patterns.
Larger unit sizes mean that users have to upload larger amounts of data when updating any part of the file.
Smaller unit sizes result in smaller upload times for small updates; small unit sizes, however, incur considerable overhead when a user updates significant parts of the file (i.e., when the update affects a large number of units).
In Section~\ref{subsec:evaluation2}, we study the performance of writing and reading multiple units.

\begin{table}[t]
\centering
\scalebox{1.5}{\begin{tabular}{|l|c|}
  \hline
   & \textbf{Peak Throughput (Mbps)}\\
   \hline
 \textbf{Write}  & 43.39\\
  \hline
 \textbf{Read}  & 29.52\\
 \hline
\end{tabular}}
\vspace{5 pt}
\caption{Peak throughput exhibited by our implementation. We assume the default parameters in Table~\ref{tab:nominal}. Each data point is averaged over 20 different measurements.}
\label{tab:throughput}
\end{table}

\subsection{Multiple units}\label{subsec:evaluation2}

After analyzing the performance of writing/reading a single file unit in \comrade, we now proceed to assess the performance of writing/reading
multiple units of a given file.
This corresponds to the typical case where a user writes/reads a file comprised of multiple units, or updates multiple units of the file.

For that purpose, we measure the peak throughput exhibited by our prototype implementation.
This is achieved as follows.
Given the default parameters (cf. Table~\ref{tab:nominal}), we increase the number of units that are concurrently
written/read to/from Amazon S3 until the throughput is saturated.
We then compute the peak throughput as the maximum aggregated amount of data (in bits) that can be written/read per second to/from Amazon S3.
Our results (cf. Table~\ref{tab:throughput}) show that the peak throughput is above 29 Mbps for both $\mathrm{write}$ and $\mathrm{read}$ operations.

In summary, we conclude that \comrade{}'s overhead can be well tolerated in
user-driven (i.e., highly interactive) scenarios
where users author files and upload/download content to the cloud.
We further point out that in a practical use case, users bear
the full cost of file write/read only once.
Since \comrade{} supports versioning, once a user has uploaded/downloaded the entire file, subsequent updates can be made to individual units in order to maximize the performance of the scheme.

\section{Discussion}\label{sec:disc}

In this section, we discuss further insights with respect to the design of \comrade.

\noindent \textbf{Transparency to Users}\mbox{}

As explained earlier, \comrade{} enables users to coordinate access control to online content in a distributed manner.
We stress that all the operations in \comrade{} are implemented by the client application described in Section~\ref{sec:implementation}.
Users need not ``manually'' distribute or fetch tokens.
In fact, users are only required to set the list of owners for the files they create and to define the access policy on the files for which they are appointed as owners.

\noindent \textbf{Changing threshold $t$}\mbox{}

To maintain consistency in \comrade, we do not support the change of threshold $t$ for any file $\F$.
More specifically, if an owner wants to change the threshold, say from $t$ to $t'$, he must compute and distribute new tokens to \emph{all} owners in $\cal{O}$.
That is, \emph{all} owners in $\cal{O}$ must replace their older tokens with the newly received ones.
However, since each user has full rights on the tokens he handles, there is no mechanism to force all owners to accept these changes, and replace their tokens.
This can lead to an inconsistent state in which some tokens correspond to one file version (with threshold $t$),
while other tokens correspond to another version (with threshold $t'$). 
This is why \comrade{} does not support changing the threshold.

\noindent \textbf{Adding/Revoking Owners}\mbox{}

Our model assumes that the set of owners $\cal O$ is defined before file creation.
Adding an owner in \comrade{} requires that either the original file creator or at least $t$ out of the $n$ owners to provide the new owner with his set of tokens. However, revoking ownership rights from an owner, say $O_j$, may not be feasible since tokens cannot be removed from $O_j$'s storage on \server{} without his consent. One possible solution would be to re-encode the file and distribute new tokens to owners in $\mathcal{O}\setminus\{O_j\}$.
Nevertheless, similar to the case of changing the threshold $t$, some of the owners in $\mathcal{O}\setminus\{O_j\}$ may decide to discard the new tokens
and keep the old ones---leading to an inconsistent state.

\noindent \textbf{Fine-Grained Per-Version Access Control}\mbox{}

\comrade{} enables owners to perform per-version access control.
That is, owner $O_j$ can, for example, grant to $U_l$ \rpermit\ access to version $x$ of a file $\F$ and deny to the same user access to version $x'$ of the same file.
This can be useful in collaborative scenarios where some versions of a given file may contain information only intended for a subset of the users (e.g., due to IPR protection).

Note that, due to versioning, a given unit may span several versions of file $\F$.
For example, Figure~\ref{fig:version} shows that version 3 and 4 share the same unit $\F_1$.
Nevertheless, this is transparent to the user who only decides whether to grant/deny access to a given version $x$, while tokens are handled by the client application.
In particular granting/denying \rpermit\ access rights to version $x$ of file $F$ is achieved as follows:

\begin{description}
\item[\textbf{Granting read access to version $x$}]\mbox{}\\
The client endorses and grants \rpermit\ access to the \emph{most recent version} of tokens pertaining to units that have been updated up to (and including) version $x$.
\item[\textbf{Denying read access to version $x$}]\mbox{}\\
The client only denies \rpermit\ access to the tokens that were only uploaded at the time when version $x$ was created, and were subsequently updated in any later version of the file for which $O_j$ wants $U_j$ to grant access to. 
\end{description}

\section{Related Work}\label{sec:related}
To the best of our knowledge, this is the first work to (i) formulate and solve
the problem of distributed enforcement of shared ownership policies.
In the following we survey relevant related work in the areas of information dispersal, all-or-nothing transformations, and access control.

\noindent \textbf{Secret Sharing and Information Dispersal}\mbox{}

Secret sharing schemes~\cite{beimel11iwcc} allow a dealer to distribute a secret among a number of shareholders, such that only authorized subsets of shareholders
can reconstruct the secret.
In threshold secret sharing schemes~\cite{shamir79cacm,charnes94ccs}, the dealer defines a threshold $t$
and each set of shareholders of cardinality equal to or greater than $t$ is authorized to reconstruct the secret.
Secret sharing guarantees security (i.e., the secret cannot be recovered) against a non-authorized subset of shareholders;
however, they incur a high computation/storage cost, which makes them impractical for sharing large files.

Rabin~\cite{Rabin} proposed an information dispersal algorithm with smaller overhead than the
one of~\cite{shamir79cacm}, however, his proposal does not
provide any security guarantees when a small number of shares (less than the reconstruction threshold) are available.
Krawczyk~\cite{kraw} proposed to combine both Shamir's~\cite{shamir79cacm} and Rabin's~\cite{Rabin} approaches;
in \cite{kraw} a file is first encrypted using AES and then dispersed using the scheme in~\cite{Rabin},
while the encryption key is shared using the scheme in~\cite{shamir79cacm}.

Information dispersal based on erasure codes~\cite{lint82book} has
been proven as an effective tool to provide reliability in a number of
cloud-based storage systems \cite{xia07sc,abd-el-malek05sosp,kubiatowicz00asplos,aguilera05dsn}.
Erasure codes enable users to distribute their data on a number of
servers and recover it despite some servers failures.

Ramp schemes \cite{blakley84crypto} constitute a trade-off between the security guarantees of
secret sharing and the efficiency of information dispersal algorithms.
A ramp scheme achieves higher ``code rates'' than secret sharing and
features two thresholds $t_1,t_2$. At least $t_2$ shares are required to reconstruct the
secret and less than $t_1$ shares provide no information about the secret;
a number of shares between $t_1$ and $t_2$ leak ``some'' information.

\noindent \textbf{All or Nothing Transformations}\mbox{}
All-or-nothing transformations (AONTs)
were first introduced in~\cite{rivest97fse} and later investigated in \cite{Boyko,desai00crypto}.
The majority of AONTs leverage a secret key that is embedded in the output blocks.
Once all output blocks are available, the key can be recovered and single blocks can be reverted.

Rivest~\cite{rivest97fse} also mentioned a transformation that is inspired by Fast Fourier Transform.
Van Dijk et al.~\cite{vandijk12ccs} later on leveraged Rivest's transformation to construct a ``proof of encryption'' of files in the cloud.
In this paper, we extend the use of Rivest's transformation to construct an AONT scheme, \aont, that supports the distributed enforcement of shared ownership in the cloud.

Resch et al.~\cite{resch11fast} combine AONT and information dispersal
to provide both fault-tolerance (i.e., decoding requires only $t$ out of $n$ shares) and data secrecy (i.e., confidentiality
is guaranteed w.r.t. parties that collect less than $t$ shares), in the context of distributed storage systems.
In~\cite{resch11fast}, however, an adversary who caches the encryption key can still decode single shares.

\noindent \textbf{Access Control Systems}\mbox{}

Current state-of-the-art access control systems---such as SecPAL~\cite{SecPAL}, KeyNote~\cite{KeyNote},
and Delegation Logic~\cite{DL}---can in principle express
$t$ out of $n$ policies. These languages, however, rely on the
presence of a centralized PDP component to evaluate their policies.
Furthermore, their PDPs cannot be deployed within a third-party cloud platform.
As explained in Section~\ref{sec:model}, these access control systems rely on an administrator to define and modify
access control policies. In our setting, this means that a set of owners has to
elect one enforcer who has unilateral powers over their files.

\section{Conclusion}\label{sec:conclusion}

Even though existing cloud platforms are used as collaborative
platforms, they surprisingly do not support any notion of shared ownership.
We consider this to be a severe limitation because
collaborators cannot jointly decide how their resources are used.
The problem of enforcing shared ownership in the
cloud is even more difficult since a cloud platform does not
allow deployment of a third-party enforcement component.

In this paper, we introduced a novel concept of shared ownership and
we described it through a formal access control model, called
\dofac{}. We also proposed our scheme, \comrade{}, that distributively
enforces \dofac. \comrade{} can be used in existing clouds without
requiring any modifications to the platforms. We implemented and
evaluated the performance of our solution within Amazon S3. Our
results show that \comrade{} scales well with the file size and with
the number of users.
%

\bibliographystyle{abbrv}
\bibliography{biblio}

\section*{Appendix}

\subsection{Datalog Encoding of the \dofac{} Access Control Policy}
\label{datalog}

To show that \dofac{} can be centrally enforced, we present its
encoding in the Datalog logic-programming language. We chose Datalog
because it is the basis for implementing most of the existing
credential-based access control systems (e.g., Trust Management systems
such as SecPAL~\cite{SecPAL}, DL~\cite{DL}, etc.).

We represent a file system state as a Datalog database \cite{Datalog}
that has a set of relations describing each file's owners and its
threshold. The \dofac{} access control policy is represented as a set of
Datalog rules (clauses). We translate requests and credentials into
Datalog clauses, which are then evaluated over the current state
together with the SOM policy.

We first give a brief overview of Datalog (see \cite{Datalog} for a
more extensive survey). A Datalog program is a
finite set of clauses of the form:
$
S \leftarrow L_1,\dots,L_n\ ,
$
where $S$ and $L_i$ are function-free first-order literals of the form
$predicate(arg_1,\dots,\arg_n)$. We refer to $S$ as the head of the
clause, and to $L_i$ as a body literal. We adopt the following
notation: a variable starts with the $?$ character, a constant starts
with a capital letter, and a predicate name starts with a lower-case
letter.
A clause with no body literals is called a \emph{fact}. All clauses
are safe: all variables that appear in a head literal also appear in
at least one body literal. A Datalog program can be split into two
sets of clauses: \emph{EDB} and \emph{IDB}. \emph{EDB} is a set of
facts whose head literals do not appear as head literals in any other
clause.  All other clauses are in the \emph{IDB} set. Intuitively, we
think of an \emph{EDB} as an input for computing all implied facts by
the clauses in the \emph{IDB} set. The declarative semantics of a
Datalog program interpret each clause as a first-order sentence:
$\forall \bar{x} L_1 \wedge \dots \wedge L_i \rightarrow S$, and take
a whole program to be a conjunction of its clauses. For each
program $\mathcal{P} = \textit{IDB} \cup \textit{EDB}$, let
$\sigma(\textit{IDB}, \textit{EDB}) = \{atom\ |\
\mathcal{I}(\mathcal{P}) \models atom\}$, where
$\mathcal{I}(\mathcal{P})$ is the first-order translation of
$\mathcal{P}$, and $\models$ is the logical implication.

A \dofac{} state $s$ is a tuple $(\textit{Files}, \textit{Users},
\textit{Owns}, \textit{Thresholds})$ where $\textit{Files}$ denotes a set of
strings representing file names, $Users$ is a set of users, $\textit{Owns}$ is
a subset of $2^{\textit{Users} \times \textit{Files}}$, and
$\textit{Thresholds}$ is a mapping from $\textit{Files}$ into $\mathbb{N}$.
For a state $s$, we define a set $EDB_s$ containing
all ground atoms: $\textit{file}(File)$,
$\textit{user}(\textit{User})$, $\textit{owns}(\textit{User},
\textit{File})$, and $\mathit{threshold}(\textit{File},N)$.
A request $r$ is a tuple $(R, \mathcal{C})$, where $R$ is a request
credential
submitted by a user, and $\mathcal{C}$ is a set of
available credentials. Credentials can be either submitted by a user,
or kept in a separate storage and appended to each request.
Given a request, $\mathcal{T}(C)$ generates the following set of
Datalog rules $IDB_r$:
\begin{align*}
\mathcal{T}(O\ \textbf{says}\ U\ \textbf{can}\
\mathit{actionOp}(\F)) &= says(O, U, \mathit{actionOp}, F)
\end{align*}

The translation of $R$ is similar, except that we do not generate $says$ facts but
Datalog queries:
\begin{align*}
\mathcal{T}(U\ \textbf{reqs}\ \mathit{actionOp}(\F) &=  can(U, \mathit{actionOp}, F)
\end{align*}

The set $\mathcal{A}_s$ is a set of Datalog rules
\emph{parameterized} on $s$, that enforces the shared ownership:
\begin{align*}
can&(?U, \mathit{actionOp}, ?F) \leftarrow file(?F), user(?U),\\
&threshold(?F, ?T),\\
&\textbf{[[}says(?O_1, ?U,\mathit{actionOp}, ?F), \dots,\\
&\quad says(?O_{?T}, ?U,\mathit{actionOp}, F),\\
&\ \ owns(?O_1, ?F),\dots,owns(?O_{T}, ?F),\\
&\ \ ?O_1 \neq ?O_2, \dots, ?O_1 \neq ?O_{?T},\dots,?O_{?T-1} \neq
?O_{?T}\textbf{]]}
\end{align*}

Intuitively, the given rule is a \emph{template} rule that
instantiates the necessary clauses for all $actionOp$ operations. The
variable $?T$ denotes a threshold, and $?U$ denotes a user. The reason
for doing so is to correctly enforce the current (for the given state
$s$) threshold $t$ for a particular file.  In short, we need to
generate the correct number of $?O_i$ variables for each file and its
threshold in $s$. To represent this \emph{dynamic} part of a clause
(that is dynamically adjusted for each state), we enclose it within
$\textbf{[[}$ and $\textbf{]]}$ brackets.

Finally, given a \dofac{} state $s$, and a request $(R, \mathcal{C})$, we
say that \dofac{}'s Datalog-based Policy Decision Point (PDP) grants $R$ if and only if:
$
\mathcal{I}(EDB_s \cup \mathcal{A}_s \cup \mathcal{T}(\mathcal{C})) \models
\mathcal{T}(R)
$.

\subsection{Security Analysis of SFD}

We treat the information dispersal encoder of $\mathrm{SFD.Encode}(\cdot)$ as a ramp scheme~\cite{blakley84crypto}.
In particular, to maximize the code rate, we assume a $(0,t,n)$-ramp scheme, with $t$ as reconstruction threshold and $0$ as privacy threshold\footnote{This means that any single chunk leaks information about the encoded input.}.
Let $\bar{F}$ denote the input to the information dispersal encoder.
Hence, at least $t$ chunks are necessary to reconstruct $\bar{F}$; however, each single chunk leaks some information about $\bar{F}$.
More specifically, if the adversary $\cal A$ is given $l$ out of $n$ chunks outputted by $\mathrm{SFD.Encode}(t,n,F,K,\lambda)$, then the following holds:
\[
H(\bar{F}|{\cal A})=
\left\{
  \begin{array}{l l}
    \frac{l}{t}H(\bar{F})   & \quad \textnormal{if $l<t$}\\
    0                       & \quad \textnormal{if $l\geq t$}
  \end{array}
\right.
\]
\noindent where $H(\cdot)$ denotes the entropy.\\

Since $\bar{F}$ is the output of $\mathrm{AONT}$ (which, is essentially a block cipher) we can assume $H(\bar{F})=m\lambda$.

Therefore, if $\cal A$ is given $l=t-1$ chunks, the uncertainty on $\bar{F}$ is reduced by a factor of $t$; that is, $H(\bar{F}|{\cal A})=\frac{m}{t}\lambda$.
Without loss of generality, we can assume that when $m=t$ and given $t-1$ SFD encoded chunks, $\cal A$ can decode all but one block of the $\mathrm{AONT}$ output.

Hence, we are left to show that, given $K$ and $\bar{f}_{i_1},\ldots,\bar{f}_{i_{m-1}}$, $\cal A$ has negligible advantage in recovering $f\subseteq F$, i.e., a substring of $F$ of size $\lambda$.
We start noting that the security of the underlying block-cipher prevents the adversary from recovering partial bits of any cleartext block.
That is, the adversary can only learn entire blocks of cleartext. Therefore, we focus our analysis on an adversary that tries to recover any cleartext block $f_i\in F$.
Furthermore, in our definition the adversary is not allowed to store any blocks of ciphertext/cleartext nor is she allowed to store any intermediate block produced by \aont.
Indeed, we argue that if a rationale adversary were to store any block of data, then she would store blocks of cleartext (i.e., the actual file) in order to easily access it despite revocation.
No security mechanism can cater for effective revocation if the adversary has a local copy of the protected resource.

We prove the security of \aont{} by induction.
Let \aont{}(4) be the graph that defines the operations of \aont{} when $m=4$.
Clearly, recovering any input block of \aont{}(4) requires all 4 output blocks.
Recall that each output block is $\lambda$-bit in size and has high-entropy (since it is the output of a block-cipher).
Therefore, given all but one output block, $\cal A$ has only negligible advantage in guessing the missing block and recovering any input block.

Similarly, an \aont{}(8) graph has two \aont{}(4) subgraphs (one left and one right), plus an additional round of encryption.
Recovering any input block of \aont{}(8) requires one input blocks of each of the two \aont{}(4) subgraphs which, in turn, require all their output blocks.
Therefore, recovering any input block of \aont{}(8) requires all 8 output blocks.
By iterating this analysis for larger graph sizes, we can easily prove that for any $m$,
recovering any input block of \aont{}($2^m$) requires all $2^m$ output blocks.

\subsection{Security Analysis of CRSS}

We now prove the security of our CRSS scheme (Section~\ref{subsec:CRSS}). More specifically, we show that a p.p.t. algorithm $\cal A$ that has non-negligible advantage in winning \texttt{Game Priv*} defined in Section~\ref{subsec:CRSS},
can be used by p.p.t. algorithm $\cal B$ as an internal routine to break the security of the threshold secret sharing scheme by Charnes et al.~\cite{charnes94ccs}.

That is, on one side $\cal B$ is challenged by a challenger $C$ to find the secret $g^x$ given only $t-1$ shares as output by $g^x,\{x_1,\ldots,x_n\}\leftarrow \mathrm{SS.Share}(-,t,n)$.
On the other side, $\cal B$ challenges $\cal A$ to break the security of CRSS.\\
Simulation starts with $\cal A$ who submits two messages $x_0,x_1$ of equal length.
Those are forwarded by $\cal B$ to $C$ that flips an unbiased coin $b$ and secret-shares $x_b$. Given our particular secret sharing scheme, we assume that the secret to be reconstructed is $g^{x_b}$. At this time, $\cal B$ also picks a random $t-1$ polynomial $Y$, such that $Y(0)=0$ and computes $y_i=Y(i)$, for $1\leq i\leq n$.

During Type-1 queries, $\cal A$ submits indexes $i_1,\ldots,i_{t'}$.
$\cal B$ forwards them to $C$ that replies with $\{x_{i_1},\ldots,x_{i_{t'}}\}$.
$\cal B$ then sends to $\cal A$ shares $\{s_{i_1},\ldots,s_{i_{t'}}\}$ where $s_{i_l}=(x_{i_l},y_{i_l})$, for $1\leq l \leq t'$.

Similarity, during a Type-2 query $\cal A$ submits a fresh identity $U_j$ and $t''$ indexes $i'_1,\ldots,i'_{t''}$.
For each identity $U_j$, $\cal B$ picks at random $h_j\in Z_q$ and sets $H(U_j)=g^{h_j}$.
If an index $i'_l$ has been submitted during type-1 query, $\cal B$ knows $x_{i'_l}$ and can compute the delegation $d_{i'_l,j}=g^x_{i'_l}H(U_j)^{y_{i'_l}}$.
Otherwise, $\cal B$ asks $C$ for share $x_{i'_l}$ and computes the corresponding delegation.

Note that $\cal A$ can submit a Type-2 query if only if $t'<t-1$. Therefore $\cal B$ can still ask $C$ for the missing shares.
During the guess stage, $\cal A$ will output his guess $b'$ and $\cal B$ will use it as its own guess towards $C$.

Since $H(\cdot)$ is a random oracle and $\cal A$ has a non-negligible advantage in guessing $b$, then $B$ has the same advantage in
breaking the security of~\cite{charnes94ccs}, thus concluding our analysis.

\end{document}